\documentclass{phb-proc4-auth}

\usepackage{graphicx}
\usepackage{amssymb}
\usepackage{amsmath}
\usepackage{amsfonts}

\begin{document}
\begin{frontmatter}


\journal{SCES '04}


\title{Critical theories for the pseudogap Kondo problem}
\author{Lars Fritz, Marijana Kir\'{c}an, and Matthias Vojta}
\address{
Institut f\"ur Theorie der Kondensierten Materie,
Universit\"at Karlsruhe, 76128 Karlsruhe, Germany
}

%
%
%
%


%
%
%
%



\begin{abstract}
We discuss quantum phase transitions in the pseudogap Kondo
problem, which describes a magnetic moment coupled to conduction electrons
with a power-law density of states, $\rho(\omega)\propto|\omega|^r$.
We show that different perturbative expansions, together with renormalization
group techniques, provide effective low-energy field theories for the
relevant critical fixed points.
In particular, we review expansions near the lower-critical and
upper-critical dimensions of the problem, being $r=0$ and $r=1$, respectively.
\end{abstract}

%
%



\end{frontmatter}

%
%
%
%
%

The pseudogap Kondo problem \cite{withoff,GBI,insi}
has attracted substantial attention in recent years, due to the presence of a non-trivial
quantum phase transition between an unscreened and a screened moment as
function of the Kondo coupling.
This phase transition, characterized by universal local-moment fluctuations
and local non-Fermi liquid behavior, is a paradigmatic example for local
quantum criticality.

The problem can be formulated in the language of either a Kondo or an Anderson impurity
model:
\begin{align}
  H_{\text{K}}&=J_{\text{K}} \mathbf{S}\cdot \mathbf{s}_0+H_{\text{b}},\\
  H_{\text{A}}&=\epsilon_0 f^{\dagger}_{\sigma} f_{\sigma}+U_0 n_{f_{\uparrow}} n_{f_{\downarrow}}
               +V_0(f_{\sigma}^{\dagger}c_{0\sigma}+\text{h.c.})+H_{\text{b}},\nonumber
\end{align}
where $J_{\text{K}}$ is the Kondo coupling, $\mathbf{S}$ the impurity spin, $\mathbf{s}_0$
the spin of the conduction electrons at the impurity site,
$\epsilon_0$ the impurity energy, $V_0$ the hybridization
and $U_0$ the on-site repulsion.
The fermionic bath is represented by
\begin{align}
  H_{\text{b}}&=\int_{-\Lambda}^{\Lambda} \text{d}k |k|^{r} k\, c^{\dagger}_{k \sigma} c_{k \sigma},
\end{align}
with a density of states following a power law at low energies,
$\rho(\omega) \propto |\omega|^r$ ($r\geq 0$).
The above models can, e.g., describe impurity moments in
unconventional superconductors or in semimetals.

Numerical work \cite{GBI} has mapped out the phase diagrams of the above models and
uncovered a rich fixed point structure, which depends on the bath exponent $r$
and the presence/absence of particle-hole (p-h) symmetry.
Recent analytical progress \cite{MKMV,MVLF}
allowed to identify the low-energy effective field theories
for both the p-h symmetric and asymmetric critical fixed points.
Three different $\epsilon$-type expansions, with the small parameters being $r$,
$(1/2\!-\!r)$, and $(1\!-\!r)$, can be used to determine the universal critical
properties.
Interestingly, $r=0$ and $r=1$ play the roles of the lower-critical and upper-critical
dimensions of the problem, implying that strong hyperscaling properties are obeyed near
the phase transition for $0<r<1$.

We briefly summarize the main results.
In the \emph{p-h symmetric case} a phase transition occurs at a critical $J_{\rm c}$
for $0<r<1/2$ only.
For small $r$, this transition can be accessed from the weak-coupling limit
of the Kondo model, using a generalization of Anderson's poor man's scaling.
The flow of the renormalized Kondo coupling $j$ is given by the beta function
\begin{equation}
\label{betaj}
\beta(j) = r j - j^2 \,,
\end{equation}
and an unstable fixed point exists at $j^\ast = r$.
We see that the Kondo coupling is marginal at the lower-critical dimension $r=0$,
and the expansion is controlled by the small parameter $r$.
For $r$ near $1/2$ the p-h symmetric critical fixed point moves
to strong Kondo coupling,
and the language of the p-h symmetric Anderson model \cite{MVLF} becomes
more appropriate.
Expanding in the renormalized hybridization $v$ and on-site interaction $u$,
the flow equations
\begin{align}
\label{betau}
&\beta(v) = - \frac{1-r}{2} \, v +  v^3 \,, \nonumber\\
&\beta(u) = (1-2r)\, u -  \frac{3(\pi-2 \ln 4)}{\pi^2}\, u^3
\end{align}
capture various fixed points;
notably the RG equation for $v$ is {\em exact} to all orders \cite{MVLF}.
For $r<1/2$ a stable fixed point is at
${v^\ast}^2 = (1-r) / 2$, $u^\ast=0$, which represents nothing but the physics of the
non-interacting resonant level model with a power-law density of states --
interestingly, the impurity spin is not fully screened for $r>0$, and
the residual entropy is $2r\ln 2$.
This fixed point is identical with the symmetric strong-coupling fixed point
of Gonzalez-Buxton and Ingersent\cite{GBI}; it becomes unstable for
$r>1/2$.
Additionally, for $r<1/2$ there is a pair of critical fixed points located at
${v^\ast}^2 = (1-r) / 2$,
${u^\ast}^2 = \pi^2 (1-2r) / [3(\pi-2\ln 4)]$.
These describe the transition between an unscreened (spin or charge) moment phase and
the symmetric strong-coupling (i.e. screened) phase.
Importantly, both (\ref{betaj}) and (\ref{betau}) capture the same critical
fixed point.
Using (\ref{betau}), the expansion is performed around the strong-coupling fixed point,
and is valid for $u^\ast\ll 1$, i.e., for $1/2-r \ll 1$.

In the \emph{p-h asymmetric case} a phase transition occurs for all $r>0$;
here the strong-coupling phase is maximally p-h asymmetric and corresponds to a fully
screened spin for all $r$.
Remarkably, for $0<r< r^{\star}\approx 0.375$ p-h symmetry is dynamically restored
at the phase transition, and the critical properties are described by the expansions discussed
above.
For $r>r^{\star}$ there exists an additional critical fixed point with finite p-h asymmetry,
and its properties are best discussed using an infinite-$U_0$ Anderson model \cite{MVLF}.
Expanding around the $v=0$, $\epsilon=0$ limit ($\epsilon$: renormalized on-site energy),
one finds the flow equations
\begin{align}
&\beta(v) = - \frac{1-r}{2} v + \frac{3}{2} v^3 \,, \nonumber \\
&\beta(\epsilon) = - \epsilon + 3 v^2 \epsilon - v^2 \,,
\label{beta}
\end{align}
with a critical fixed point
${v^\ast}^2 = (1-r)/3$, $\epsilon^\ast = - (1-r)/3$ for $r<1$, and
$v^\ast=\epsilon^\ast=0$ for $r>1$.
Note that the flow is very similar to the one of a $\phi^4$ model,
with $(1-r)$ playing the role of $(4-d)$, as
the Anderson model hybridization is marginal at $r=1$.
Clearly, the critical fixed point is interacting for $r<1$ (the analogue of the
Wilson-Fisher fixed point), whereas for $r>1$ we have a level crossing with
perturbative corrections (the analogue of the Gaussian fixed point).
This justifies to identify $r=1$ with the upper-critical dimension of the
problem.
RG analysis has shown that all fixed points presented above are stable
with respect to spin anisotropies.

\begin{figure}
\begin{center}
\includegraphics[width=7.5 cm]{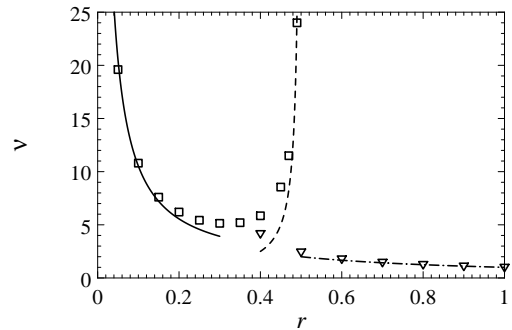}
\caption{
NRG data for the correlation length exponent $\nu$
at the symmetric (squares) and asymmetric (triangles) critical points
together with analytical results from the three perturbative RG expansions
described in the text.
}
\label{fig_nu}
\end{center}
\end{figure}

We have used these three $\epsilon$-type expansions together with renormalized perturbation
theory to determine the critical properties near the quantum phase transitions,
such as the impurity and local susceptibilities, the impurity entropy, and
$T$-matrix \cite{MKMV,MVLF}.
In Fig. \ref{fig_nu} the correlation length exponent $\nu$ obtained by perturbative
RG is compared with NRG results.
This exponent describes the vanishing of the characteristic energy scale $T^\star$,
above which critical behaviour occurs,
upon tuning the system through the transition: $T^\ast \propto |J_{\rm K}-J_{\rm c}|^\nu$.
At the symmetric critical fixed point $\nu$ diverges for both $r\to 0^+$
and $r\to\frac{1}{2}^-$.

Let us briefly mention our result for the conduction electron $T$-matrix,
describing the scattering of the $c$ electrons off the impurity.
At criticality it follows a power law
$
T(\omega)\propto 1/\omega^{1-\eta_T}
$.
One finds the {\em exact} result $\eta_T=1-r$ for $r<1$, i.e., for all interacting fixed points
considered above the $T$-matrix follows $T(\omega)\propto\omega^{-r}$.
This perfectly agrees with NRG calculations.
Furthermore, at the upper-critical dimension
we find $\Im\text{m}\, T(\omega)\propto\frac{1}{\omega |\log{\omega}|^{2}}$ --
this applies, e.g., to a Kondo impurity in a $d$-wave superconductor.

To summarize, we have analyzed quantum phase transitions in the pseudogap Kondo problem
using different RG expansions. We have identified the relevant degrees of freedom
at the corresponding critical fixed points which allow us to perform expansions
around marginally relevant operators in the respective $r$-range.


%
%
%
%


\end{document}